\documentclass{article}
\usepackage{amsmath,amsfonts}
\newcommand{\E}{\mathrm{e}}
\newcommand{\pd}{\partial}
\newcommand{\dd}{\mathrm{d}}
\newcommand\hR{\hat R}
\newcommand{\I}{\mathrm i}
\newcommand{\s}{_{\mathrm{sp}}}
\def\ket#1{|#1\rangle}

\newcommand\hH{\hat H}
\newcommand\vt\vartheta
\newcommand\vr\varrho
\newcommand\eps\epsilon
\newcommand\hvr{\hat\varrho}
\newcommand\hr{\hat\rho}
\newcommand\ZZ{\mathbb Z}
\newcommand\ex{^{\mathrm{exact}}}
\newcommand\beq{\begin{equation}}
\newcommand\eeq{\end{equation}}
\newcommand{\st}{{\vartheta}}
\newcommand{\half}{\frac{1}{2}}
 \DeclareMathOperator\Res{Res}
\DeclareMathOperator\Real{Re} \DeclareMathOperator\Imag{Im}
\DeclareMathOperator\sgn{sgn}
\newcommand\FT{_{_\mathrm{FT}}}
\newcommand\F{_{\mathrm F}}
\begin{document}

\begin{titlepage}

May 2006
\hfill CCNY-HEP-06/6
\vskip 1cm
\begin{center}
\renewcommand{\thefootnote}{\fnsymbol{footnote}}
{\LARGE The fermion density operator in the droplet bosonization picture
} \vskip .45in Alberto Enciso$^{a}$ and Alexios P. Polychronakos$^{b}$ \vskip
.45in

{\em $^{a}$Departamento de F\'\i sica Te\'orica II, Universidad Complutense \\
        28040 Madrid, Spain \\
        {\rm {\small aenciso@fis.ucm.es}}
\vskip .1in
        $^{b}$Physics Department, City College of the CUNY \\
        New York, NY 10031, USA \\
        {\rm {\small alexios@sci.ccny.cuny.edu}}
}
\end{center}
\vskip 0.45in
\begin{abstract}
We derive the phase space particle density operator in the `droplet' picture
of bosonization in terms of the boundary operator. We demonstrate that it
satisfies the correct algebra and acts on the proper Hilbert space
describing the underlying fermion system, and therefore it can be used
to bosonize any hamiltonian or related operator. As a demonstration we show
that it reproduces the correct excitation energies for a system of free fermions
with arbitrary dispersion relations.
\end{abstract}

\end{titlepage}

\newpage
\renewcommand{\thepage}{\arabic{page}}

\setcounter{page}{1}
\setcounter{footnote}{0}

\section{Introduction}

The description of a one-dimensional fermion system in terms of bosonic degrees
of freedom, called bosonization, is a well-established and useful technique
\cite{Early,Bozo,WZW}.
The advantage of such a transmutation lies
in the fact that the dynamics of the equivalent bosonic theory
encodes many-body excitations of the fermion system, with large coherent (classical)
bosonic excitations representing large collective states of the fermions.
It provides, therefore, a convenient setting to study strongly correlated fermions.

It is intuitively obvious that such bosonic descriptions must trivially exist
in any number of dimensions,
at least within each fermion number sector: any excitation of the system
that does not change the number of fermions (or changes them by an even number)
is inherently bosonic and can be obtained with the action of a bosonic operator
on the ground state.
The full theory, therefore, can be described in terms of a collection of these
bosonic operators. Examples of such operators are the ``hydrodynamic'' current
and particle densities, or the more general phase space density operator that
will be examined in this paper.

Such descriptions, however, do not qualify as proper bosonizations. The
set of bosonic operators describing the system would form, in general, an
overcomplete set and would make the identification of the proper features
of the fermion system difficult. More formally, the algebra of the bosonic operators
would, in general, admit several irreducible representations, of which only one
would be the fermion Hilbert space. Consequently, the classical action for the
bosonic fields could not be used to identify proper collective excitations
of the fermions.

The above-mentioned hydrodynamic or phase space density variables
provide examples of such a redundancy: their algebra is the same for
all systems, and admits representations that describe fermions, bosons, as well
as a host of other systems (parastatistics, fractional filling states etc.)
A successful bosonization of a system should realize the
fermion states as its {\it complete, irreducible} Hilbert space.

Bosonization works successfully for one spatial dimension, where special
properties of line-like spaces come into play. In particular, statistics and spin
are essentially irrelevant notions in one dimension and thus fermionic and
bosonic theories could a priori have the same degrees of freedom. On a more
intuitive level, fermion dynamics around the Fermi (or Dirac) sea become
tantamount to one-dimensional wave propagation, with the corresponding phonon
states encoding quantum excitations \cite{Early}. Further, exponentials in the
bosonic field (`vertex operators') can be properly defined to obey fermionic
commutation relations. Overall, we have an exact mapping of states between the
two systems both at the many-body and the field theory level.

Bosonization can be motivated or formulated from various starting
points (operator, field theory, many-body etc.). A particularly intuitive
approach is to consider a semiclassical description in terms of a dense collection
of fermions forming a `droplet' in their phase space and study the dynamics
of the Fermi surface of the droplet \cite{droplet}. In one dimension, this
leads to a chiral theory that fully reproduces the standard bosonization
results. In particular, for nonrelativistic fermions, it reproduces the
results of the collective field method \cite{collect}.

The basic advantage of the phase space approach is that, in principle, it works
in any number of dimensions and it seems to capture
the relevant degrees of freedom for a successful bosonization.
In fact, an adaptation of this method properly taking into account the quantum
nature of the phase space was proposed by one of the authors as the starting
point for an {\it exact} (that is, complete and irreducible) bosonization in
arbitrary dimensions \cite{ncboz}. This could become a useful tool in the study
of correlated fermion systems in dimensions higher that one. (For other approaches
to higher dimensional bosonization see \cite{hiboz}.)

In paving the way to such applications, it is important to establish the full
description of the Fermi system in the phase space droplet picture. In
particular, it is important to have an exact mapping of all fermion operators
in terms of the (bosonic) field describing the droplet boundary. This would
provide the bosonized form of {\it any} many-body hamiltonian, including
interactions, and requires deriving the boundary field expression of a complete
set of fermion operators. Such a set is encoded by the phase space density
operator, in terms of which any other operator can be obtained.

The purpose of this paper is to do this in the controllable case of one
spatial dimension, in which standard bosonization works. Specifically,
the expression for the quantum density operator will be given as a function
of the droplet boundary field. It will be shown that this operator obeys
the correct algebra and acts on the correct Hilbert space provided by
the unique irreducible representation of the boundary field. It therefore
realizes a bosonization of any fermion system in the droplet picture
and can be used as the starting point for higher-dimensional generalizations.

\section{Review of the droplet method}

The starting point is a semiclassical description of the many-body fermion
state in terms of its phase space density. For a full discussion of this
method in any number of dimensions we refer the reader to \cite{phase}.
For our specific case of one-dimensional systems, the phase space density is
a function of two phase space variables $\rho(x,p)$ that gives the number of
particles $\dd N$ inside a small phase space area $\dd x\,\dd p$:
\[
\dd N = \rho(x,p)\, \dd x \,\dd p\,.
\]
In terms of a (large) number of individual particles, this can be though as the
smoothed-out version of the phase space point density \beq \rho(x,p) =
\sum_{i=1}^N \delta(x-x_i ) \delta (p-p_i ) \label{point} \eeq where $(x_i ,
p_i )$ are the phase space positions of the particles.

We shall consider a collection of non-interacting spinless fermions with a
single-particle hamiltonian $H\s(x,p)$. For simplicity, we assume that we have
identified Darboux coordinates in phase space, with single-particle Poisson
brackets
\[
\{ x,p \}\s = 1\,.
\]
Ordinary particle position and momentum are such coordinates, but in principle
we can use any other canonical pair.

Semiclassically there is one quantum state per area $2\pi\hbar$
in phase space. The exclusion principle mandates that we can put
at most one particle per state, and implies a
maximal value for the fermion density $\rho_o = 1/2\pi\hbar$,
corresponding to a dense collection of fermions each occupying one
quantum of phase space area.

The ground state of the system would consist of fermions occupying all
available single-particle states up to a maximal energy $E\F$. Semiclassically,
this corresponds to a density $\rho = \rho_o$ for all points in phase space
with energy $E = H\s (x,p) < E\F$ and $\rho=0$ for points with $E = H\s(x,p)
> E\F$. This is the `phase space droplet' picture. This state is parametrized
by its one-dimensional boundary, in this case the set of phase space points at
energy $E\F$:
\[
\{(x_b,p_b):H\s (x_b ,p_b ) = E\F\}\,.
\]
We shall assume that this is a single connected curve, although situations
where the boundary is disconnected, corresponding to topologically nontrivial
or disconnected droplets do arise and can be easily treated. The density is
given as
\[
\rho (x,p) = \rho_o \,\st (E\F - H\s (x,p))\,,
\]
where $\st (y) = \half(1+\sgn(y))$ is the step function. The Fermi level $E\F$
and the total number of particles $N$ are related as
\[
N = \int \rho(x,p)\,\dd x\, \dd p = \frac{A\F}{2\pi\hbar}\,,
\]
where $A\F$ is the phase space area of the ground state droplet.

Excitations of the system can be visualized as deformations of the droplet, in
which its boundary deviates from the Fermi level $E\F$. The density function
remains constant and equal to $\rho_o$ inside the droplet and zero outside.
Such states are determined by the (arbitrary) shape of the boundary. A
convenient way to parametrize the boundary is to give the value of one of the
phase space coordinates, say $p$, on the boundary as a function of the other:
\[
p_b = R(x)\,.
\]
We shall call the function $R(x)$ the boundary field. The density of the
droplet is given in terms of $R(x)$ as \beq \rho(x,p) = \rho_o \st (R(x) -
p)\,. \label{rhodrop} \eeq

The phase space density description of the system has a hamiltonian structure,
inherited from the single-particle canonical structure. The density function
satisfies the field theory Poisson brackets \beq \{ \rho(x,p) , \rho(x' , p' )
\}\FT = \partial_x \rho(x,p) \delta(x-x' ) \delta' (p - p' ) - \partial_p
\rho(x,p) \delta' (x-x' ) \delta (p-p' )\,. \label{rhoPB} \eeq These are the
standard Poisson field brackets, implying the canonical structure for test
functions $\rho[f] = \int \dd x\, \dd p\, f(x,p)\, \rho(x,p)$:
\[
\{ \rho[f] , \rho[g] \}\FT = \rho[ \{ f,g \}\s ]
\]
The hamiltonian is the total energy $H$ of the state, found by summing the
energies of all particles: \beq H = \int H\s (x,p) \, \rho (x,p)\,\dd x\, \dd p
= \rho [ H\s ] \label{Hgen} \eeq The above reproduce the evolution of the field
$\rho$ as the canonical equation of motion
\[
{\dot \rho} = \{ \rho , H \}\FT = \{ \rho , H\s \}\s
\]
This is the same as the equation implied by the evolution of the particles
making up the state according to their single-particle equations of motion.

For a droplet state, the density inside the droplet remains constant, by
Liouville's theorem. Its evolution is determined by the evolution of the
boundary, that is, the boundary field $R(x)$. The hamiltonian becomes
\beq H =
\rho_o \int H\s(x,p) \, \st (R(x)-p )\,\dd x\,\dd p
\label{H} \eeq
The canonical structure for the field $R(x)$ can be determined through a
hamiltonian reduction of the original Poisson brackets (\ref{rhoPB}) to the
submanifold of functions $\rho(x,p)$ of the droplet form. The result is
\[
\{ R(x) , R(x' ) \}\FT = -\frac{1}{\rho_o} \delta' (x - x' ) = -2\pi \hbar\,
\delta' (x - x' ) \label{RPB}
\]
Conversely, the above Poisson structure for $R(x)$ implies the Poisson brackets
(\ref{rhoPB}) for the density function $\rho(x,p)$ as expressed in
(\ref{rhodrop}).

The advantage of the above semiclassical construction is that its main features
survive quantization. Specifically, the boundary field's Poisson brackets are
elevated to quantum commutators (from now on we use the convention $\hbar =1$
and denote quantum operators with hatted symbols): \beq [ \hR(x) , \hR(x' ) ] =
-2\pi \I\, \delta' (x-x' ) \label{chiral} \eeq We recognize the above
commutation relations as a chiral abelian current algebra; the field $\hR(x)$
is the chiral field that bosonizes the fermion theory around the Fermi level
$E\F$. If the fermion system has more than one Fermi points, the values of the
boundary field at each point become independent commuting chiral fields.

The above chiral algebra has a unique irreducible unitary representation. Its
states are into one-to-one correspondence with the excitations of the fermion
system around the Fermi level. Furthermore, the hamiltonian (\ref{H}),
diagonalized in the states of this representation, reproduces the
correct excitation energies of the fermion system, at least for the cases of
free relativistic or nonrelativistic fermions. We recover the results of
standard (abelian) bosonization, with the boundary field $\hR(x)$ becoming the
chiral part of the many-body particle current operator in the bosonic
description. For free fermions on a compact space there are generically two
Fermi points and the corresponding boundary fields constitute the chiral
components of the derivative of the boson field.

The mapping of states fails only when the number of excited fermions reaches
$N$, in which case we deplete the Fermi sea. This is a large-deformation effect
in the droplet picture, expressed by a condition of the form $R(x) >0$, which
would introduce nontrivial constraints in the states of the Hilbert space of
$\hR(x)$ and would complicate the picture. Such effects are purely
nonperturbatively in the number of fermions $N$ and can be ignored for large
$N$. In the sequel we shall talk about the bosonization scheme being `exact'
and will always mean up to nonperturbative effects.

\section{The algebra of density operators}\label{algebra}

The quantum density operator $\hr (x,p)$ would naively satisfy the commutation relations
implied by elevating the Poisson brackets (\ref{rhoPB}) into quantum commutators. This,
however, is not the case. To see this, start with the quantum version of the many-body
density expression (\ref{point}):
\[
\hr (x,p) = \sum_{i=1}^N \delta (x- {\hat x}_i ) \delta (p - {\hat p}_i )\,.
\]
Since ${\hat x}_i$ and ${\hat p}_i$ do not commute, the above expression is not
even hermitian and requires the specification of a quantum ordering.

It is convenient to adopt the Weyl (symmetric) ordering between ${\hat x}_i$
and ${\hat p}_i$. In terms of the Fourier transform $\hr (l,k)$ of $\hr (x,p)$
this is expressed as
\[
\hr (l,k) = \sum_{i=1}^N e^{-i l \hat x_i -i k \hat p_i}\,.
\]
Using the Baker--Campbell--Hausdorff formula, it is now a simple matter to
derive the commutators
\begin{equation}\label{sine}
[\hr(l,k),\hr(l',k')]=2\I\sin\bigg(\frac{lk'-l'k}2\bigg)\,\hr(l+l',k+k')\,.
\end{equation}
This is the famous `sine' algebra \cite{sina} that deforms the classical field
theory Poisson
brackets (\ref{rhoPB}). It is realized, for instance, by the spatial density of
planar particles in the lowest Landau level (modulo a factor arising from the
different, anti-normal ordering in that case) \cite{LLL}.
In a different basis, it is known as the $W_\infty$ algebra \cite{Wa}.
We shall simply call it the (quantum) density algebra.

The use of the above Weyl-ordered operator $\hr (x,p)$ is that, if the
single-particle quantum hamiltonian $\hH\s$ is expressed in a Weyl-ordered form
in terms of $\hat x$ and $\hat p$ (which is always possible), then the exact
quantum many-body hamiltonian can be expressed as
\[
\hH = \sum_{i=1}^N \hH\s ({\hat x}_i , {\hat p}_i ) = \int H\s (x,p) \hr
(x,p) \,\dd x\,\dd p\label{Hexact}\,,
\]
with $H\s (x,p)$ the Weyl `symbol' of $\hH\s$, that is, the Weyl-ordered
classical expression of $\hH\s$ in terms of $x$ and $p$. In effect, the
many-body hamiltonian is obtained via the (properly ordered) classical
expression by the use of $\hr(x,p)$. All other many-body operators can
similarly be obtained as appropriate expressions involving $\hr(x,p)$, which
becomes a `universal' many-body operator.

In the context of the phase space bosonization, what remains to be shown is
that the droplet expression (\ref{rhodrop}) for the quantum density operator
$\hr(x,p)$ in terms of $\hR(x)$ satisfies the proper quantum mechanical
commutation relations. If this is the case, then the bosonization of any
many-body fermion system has been achieved. Indeed, as argued before, $\hR(x)$
reproduces the proper many-body fermion Hilbert space and $\hr(x,p)$, acting on
this Hilbert space and satisfying the proper algebra, would be the correct
quantum mechanical expression for the density operator. In particular, the
expression (\ref{Hexact}) for the hamiltonian would be exact and would
reproduce the correct energy states.

It is possible to start with the expression (\ref{rhodrop}) in terms of
$\hR(x)$ and, using the operator commutation relations (\ref{chiral}) for
$\hR(x)$, show that it formally obeys the correct operator algebra \cite{Dim}.
This proof,
however, suffers from shortcomings related to the presence of singularities.
Specifically, the expression \eqref{rhodrop} for $\hr(x,p)$ is formal and
contains infinities. A proper definition requires normal ordering and the
normal-ordered expression for $\hr(x,p)$ fails to reproduce the correct
algebra.

In the next sections we show that a proper quantum mechanical definition of the
droplet density $\hr(x,p)$ requires a nontrivial modification; once this is
done, the correct operator algebra is recovered.

\section{The semiclassical droplet density operator}\label{semidrop}

Although the semiclassical droplet density presented above is not the
appropriate expression for the quantum density operator $\hr(x,p)$, it will
nevertheless be used as the starting point for an exact bosonization of the
system.

For concreteness, we consider fermions on a periodic space of period $L=2\pi$,
which allows for discrete integer Fourier modes for functions of $x$. We shall
use Darboux coordinates $(x,p)\in[-\pi,\pi]\times[0,\infty)$ where the momentum
is positive. (This amounts to considering only `half' the Fermi sea states, and
thus working around one Fermi level with one boundary field operator.) So
$(x,p)$ can be thought of as angular and radial variables describing the
droplet. The quantum boundary field $\hR(x)$ obeys the commutation relations
(\ref{chiral}), or, in terms of modes,
\begin{equation}\label{R_n}
[\hR_n,\hR_m]=n\delta_{n+m}\,,
\end{equation}
where
\[
\hR_n=\frac1{2\pi}\int\hR(x)\,\E^{-\I nx}\,\dd x\qquad (n\in\ZZ)
\]
denote the Fourier components of the field $\hR(x)$.

The above is a set of mutually commuting harmonic oscillators. In standard
current algebra representation fashion, we consider the Fock space of the
oscillators consisting on the vacuum state $\ket0$, which is annihilated by
all the positive modes $\hR_n$ ($n>0$), and all the excited states generated
by the action of creation operators $\hR_{-n}$. Since the zero mode $\hR_0$ is a
Casimir operator, its action on the vacuum $\hR_0\ket0=\frac N2\ket 0$ defines a
conserved quantity $N$, which corresponds to the number of fermions. For
concreteness, let us assume that $N$ is an odd integer. We shall use the notation
\[
\ket{n_1,\dots,n_j}=\hR_{-n_1}\cdots \hR_{-n_j}\ket0
\]
to represent the action of the negative modes; one should observe that, since
negative modes commute among themselves, any permutation of $(n_1,\dots,n_j)$
defines the same vector, so the state $\ket{n_1,\dots,n_j}$ is only defined up
to permutations.

A first attempt to define the quantum density operator would be to normal order
the expression for $\vt(\hR(x)-p)$, so as to have an operator with finite
matrix elements between excited states; that is,
\[
:\E^{\I\sum_{n\in\ZZ}c_n\hR_n}:\,=\E^{\I\sum_{n<0}c_n\hR_n}\E^{\sum_{m\geq0}c_m\hR_m}\,,
\]
which is tantamount to moving all the positive (resp.\ negative) modes to the
right (resp.\ left) in any polynomial in the Fourier components of $\hR(x)$. As
the positive and negative modes commute among themselves, the above expression
is meaningful. We can thus introduce the quantum counterpart
\begin{equation}\label{rcl}
\hvr(x,p)=\frac1{2\pi}:\vt(\hR(x)-p):
\end{equation}
of the semiclassical density.

It can be seen, however, that this is not the correct quantum expression,
as it does not satisfy the proper algebra. In fact, it is instructive to
study the spectrum of the quantum hamiltonian operator
\begin{equation}\label{H_vr}
\hH_\vr=\int H\s(x,p)\,\hvr(x,p)\,\dd x\,\dd p - E_0\,.
\end{equation}
and compare with the exact many-body fermion spectrum. Here the integral is
taken over the domain $[-\pi,\pi]\times[0,\infty)$ and we have subtracted the
inessential central term
\[
E_0=\frac1{2\pi}\int H\s(x,p)\,\vt(\hR_0-p)\,\dd x\,\dd p
\]
to shift to zero the ground state energy. Let us henceforth assume that the
single particle Hamiltonian is
\begin{equation}\label{Hsp}
H\s(x,p)=p^s
\end{equation}
for some positive integer $s$. With this assumption, one can easily find the
expression
\begin{align}\nonumber
\hH_\vr&=\frac1{2\pi(s+1)}\int:\hR(x)^{s+1}:\dd x-\frac1{s+1}\hR_0^{s+1}\\
\nonumber
&=\frac1{2\pi(s+1)}\sum_{n\in\ZZ^{s+1}}\hR_{n_1}\cdots\hR_{n_{s+1}}
\int\E^{\I(n_1+\cdots+n_{s+1})x}\dd x-\frac1{s+1}\hR_0^{s+1}\\
&=\frac1{s+1}\sum_{n\in\ZZ^{s+1}-\mathbf0,\,\sum
n_i=0}\hR_{n_1}\cdots\hR_{n_s}\label{R^s}
\end{align}
for the integral~\eqref{H_vr}. Since $\hR_0=\frac N2$, one has the series
expansion
\begin{align}\nonumber
\hH_\vr=&s\big(\tfrac
N2\big)^{s-1}\,\hR_{-n}\hR_n\\
&+\tfrac12s(s-1)\big(\tfrac N2\big)^{s-2}\,\big(\hR_{-n}\hR_{-m}\hR_{n+m}
+\hR_{-n-m}\hR_n\hR_m\big)+O(N^{s-3}\hR^4)\label{series}\,.
\end{align}
Here and in what follows, all the dummy indices $n,m$ range from 1 to $\infty$
unless otherwise stated, and summation over repeated indices is understood.

In order to study the low energy modes of $\hH_\vr$, it is convenient to define
the order of a state $\ket{n_1,\dots,n_j}$ of the Fock space to be the integer
$n_1+\cdots+n_j$. From Eq.~\eqref{R^s} it is clear that the subspace of states
of any fixed order $r$ is invariant under $\hH_\vr$. Hence both $\ket 0$
(ground state) and $\ket 1$ (first excited state) must be eigenstates of
$\hH_\vr$, whose corresponding energies
\begin{equation}\label{E_vr}
E_0^\vr=0\,,\qquad E_1^\vr=s(\tfrac N2)^{s-1}
\end{equation}
can be easily obtained from Eq.~\eqref{series}.

Let us compare energies~\eqref{E_vr} with the actual ones. The exact energies
can be obtained by quantizing the single-particle momentum, which is restricted
to take integer values, and then filling the energy levels with $N$ fermions.
Hence, the ground state energy is obtained by filling all the momenta up to
$p=\frac{N-1}2$ with fermions. The first excited level corresponds to lifting
the top fermion (at $p=\frac{N-1}2$) to the next available level, i.e.,
$p=\frac{N+1}2$. The second excitations are obtained by promoting either the
top fermion to $p=\frac{N+3}2$ or the second highest ($p=\frac{N-3}2$) to
$p=\frac{N+1}2$. Subtracting the ground state energy, this yields
the values
\begin{subequations}\label{Eex}
\begin{align}
E_0\ex&=0\,,\\\label{Eex1} \quad
E_1\ex&=2^{-s}\big[(N+1)^s-(N-1)^s\big]\,,\\\label{Eex2}
E_{2,\pm}\ex&=2^{-s}\big[\pm(N\pm3)^s\mp(N\mp1)^s\big]\,.
\end{align}
\end{subequations}
for the lowest eigenenergies of the system. Therefore, the
Hamiltonian~\eqref{H_vr} solely provides the leading semiclassical
term of the exact energy.

We therefore conclude that the normal-ordered semiclassical
density $\frac1{2\pi}\,\vt(R(x)-p)$ is, as expected, exact only in the
semiclassical limit $N\to\infty$. A further modification is needed for
the exact operator.

\section{The exact quantum density operator}\label{dropde}

The main result of this paper is the construction of a modified density
operator $\hr(x,p)$ which satisfies the sine algebra \ref{sine}
\begin{equation}
[\hr(l,k),\hr(l',k')]=2\I\sin\bigg(\frac{lk'-l'k}2\bigg)\,\hr(l+l',k+k')
\end{equation}
and provides an exact bosonization of the system via
\begin{equation}\label{Hq}
\hH=\int H\s(x,p)\,\hr(x,p)\,\dd x\,\dd p
\, .
\end{equation}
As before, $\hr(l,k)$ denotes the Fourier transform of the density operator,
in terms of which one can recover $\hr(x,p)$ as
\begin{equation}\label{hr(x,p)}
\hr(x,p)=\frac1{4\pi^2}\sum_{l=-\infty}^\infty\E^{\I lx}\int\hr(l,k)\,\E^{-\I
kp}\,\dd k
\end{equation}

We claim that the modified density operator is given by
\begin{equation}\label{hr}
\hr(l,k)=\int\frac1{4\pi\I\sin\frac k2}\,\E^{a(x,k)}\, \E^{b(x,k)}\,\E^{\I
k\hR_0}\,\E^{-\I lx}\,\dd x\,,
\end{equation}
where
\begin{equation}\label{a}
a(x,k)=\frac{2\I}n\sin\frac{nk}2\,\hR_{-n}\,\E^{-\I n x}\,,\quad
b(x,k)=\frac{2\I}n\sin\frac{nk}2\,\hR_n\,\E^{\I nx}\,.
\end{equation}
Since the classical limit amounts to $k,l \ll 1$, and the Fourier transform of
the step function is $\frac1{\I k}$, it is clear
that we recover $\hvr(x,p)$ in the classical limit from Eq.~\eqref{hr(x,p)}.

Let us prove that the modified density operator satisfies the sine
algebra~\eqref{sine}. We shall multiply each term of the Fourier series in
Eq.~\eqref{hr} by $\E^{-\frac\eps2n}$ to render the sum (strongly) convergent,
and take the limit $\eps\downarrow0$ at the end of the calculation. This is
equivalent to acting with the operators $\hr(l,k)$ on the vacuum, or states
generated from the vacuum through the action of a finite number of creation
operators $\hR_{-n}$, which is the proper physical Hilbert space. On such states
only a finite number of terms in the expansion of $\hr(l,k)$ survive and all
expressions are finite and convergent. To simplify
the notation, from now on we shall write $a,\,a'$ instead of
$a(x,k),\,a(x',k')$ (and similarly for $b$).

Since the commutator
\begin{align*}
[a,b']&=-4n\,\delta_{nm}\frac{\sin\frac{nk}2}n\frac{\sin\frac{mk}2}m
\E^{(-\I x-\frac\eps2)n}\,\E^{(\I x'-\frac\eps2)m}\\
&=\log\bigg(\frac{(1-\E^{\I(\frac{k-k'}2+x-x'+\I\eps)})(1-\E^{\I(-\frac{k-k'}2+x-x'+\I\eps)})}{(1
-\E^{\I(\frac{k+k'}2+x-x'+\I\eps)})(1-\E^{\I(-\frac{k'+k}2+x-x'+\I\eps)})}\bigg)\\
&=\log\bigg(1+\frac{2\sin\frac
k2\,\sin\frac{k'}2}{\cos\frac{k+k'}2-\cos(x-x'+\I\eps)}\bigg)
\end{align*}
is a central element, we can use the Baker--Campbell--Hausdorff formula to write
\begin{align}\nonumber
[\hr(l,k),\hr(l',k')]&=-\int\big[\E^a\E^b,\E^{a'}\E^{b'}\big]\frac{\E^{\I(k+k')\hR_0-\I(lx+l'x')}}{16\pi^2\sin\frac
k2\sin\frac{k'}2}\,\dd x\,\dd x'\\
&=\int\E^{a+a'}\E^{b+b'}\,f(x-x',k,k',l')\,\E^{\I(k+k')\hR_0-\I(l+l')x}\,\dd
x\,\dd x'\label{integral}\,,
\end{align}
where the function $f=f(x-x',k,k',l')$ is given by
\begin{align*}
f&=-\frac{\E^{\I l'(x-x')}}{16\pi^2\sin\frac
k2\,\sin\frac{k'}2}\big(\E^{[a,b']}-\E^{[a',b]}\big)\\
&=\frac{\E^{\I
l'(x-x')}}{8\pi^2}\,\bigg(\frac1{\cos\frac{k+k'}2-\cos(x-x'-\I\eps)}-
\frac1{\cos\frac{k+k'}2-\cos(x-x'+\I\eps)}\bigg)\,.
\end{align*}

To evaluate the commutator~\eqref{integral}, it is convenient to change
variables in the integral from $x$ and $x'$ to $x$ and $y=x-x'$, and use
periodicity to evaluate the $y$-integral over the boundary of the rectangle
\[
\Omega=\{y\in\mathbb C:-\pi<\Real y-\tfrac{k+k'}2<\pi,\,(\sgn l')\,\Imag
y>0\}\,.
\]
For concreteness, let us suppose that $l'$ is positive. Let us denote the
operator-valued function $\E^{a+a'}\E^{b+b'}$ by $F$ and set $K=\frac{k+k'}2$.
The function $fF$ has two poles in $\Omega$, namely $y_\pm=\pm K+\I\eps$, and
its residues are
\begin{equation}\label{Res}
\Res(fF;y_\pm)=\pm\frac{\E^{\I l'y_\pm}}{8\pi^2\sin K}\,F|_{y=y_\pm}\,.
\end{equation}
In the limit $\eps\downarrow0$, the poles tend to $\pm K$ and one can write
\begin{align}\nonumber
F|_{y=\pm K}&=\E^{\frac{2\I}n\,\hR_{-n}\E^{-\I
nx}(\sin\frac{nk}2+\sin\frac{nk'}2\,\E^{\pm\I nK})}\E^{\frac{2\I}m\hR_m\E^{\I
mx}(\sin\frac{mk}2+\sin\frac{mk'}2\,\E^{\mp\I mK})}\\ \nonumber
&=\E^{\frac{2\I}n\hR_{-n}\E^{-\I n(x\mp\frac{k'}2)}\sin nK}\E^{\frac{2\I}m\hR_m
\E^{\I m(x\mp\frac{k'}2)}\sin nK} \\
&=\E^{a(x\mp\tfrac{k'}2,k+k')}\E^{b(x\mp\tfrac{k'}2,k+k')}\,.\label{F(K)}
\end{align}
Combining Eqs.~\eqref{Res} and~\eqref{F(K)} one can evaluate the
integral~\eqref{integral} by residues as
\begin{align*}
[\hr(l,k),\hr(l',k')]&=2\pi\I\lim_{\eps\downarrow0}\int\big[\Res(fF;y_+)+\Res(fF;y_-)
\big]\,\E^{\I(k+k')\hR_0-\I(l+l')x}\,\dd x\\
&=\frac{\E^{\I l'K}}{4\pi\I}\int\frac{\E^{\I(k+k')\hR_0-\I(l+l')x}}{\sin
K}\,\E^{a(x-\tfrac{k'}2,k+k')}\E^{b(x-\tfrac{k'}2,k+k')}\,\dd x\\
&\qquad\quad-\frac{\E^{-\I
l'K}}{4\pi\I}\int\frac{\E^{\I(k+k')\hR_0-\I(l+l')x}}{\sin
K}\,\E^{a(x+\tfrac{k'}2,k+k')}\E^{b(x+\tfrac{k'}2,k+k')}\,\dd x\\
&=2\I\sin\bigg(\frac{lk'-l'k}2\bigg)\,\hr(l+l',k+k')\,.
\end{align*}
This completes the proof of Eq.~\eqref{sine}.

\section{The energy spectrum}\label{spec}

We shall now demonstrate that the spectrum of the Hamiltonian~\eqref{Hq} expressed
in terms of the above density operator encodes the
full low-energy dynamics of the $N$-fermion system.

By an appropriate canonical transformation, almost any single-particle
hamiltonian can be brought to the form
\[
H\s(x,p) = h(p)\,,
\]
where $x$ is periodic and $p>0$. This is true for hamiltonians with compact,
simply laced equal-energy lines $\{H\s = E\}$, where $x$ is the angular
variable around equal-energy lines and $p$ an appropriate conjugate variable.
Corresponding quantum hamiltonians can also be brought to this form upon proper
reordering.

It is easy to show that all such bosonized many-body hamiltonians
commute with each other. Indeed, the expression
\begin{equation}\label{h}
\hH_{_h}=\int h(p) \,\hr(x,p)\,\dd x\, \dd p
\end{equation}
will involve only the Fourier modes $\hr (0,k)$. All such modes commute with
each other, and so will the $\hH_{_h}$. They constitute a space of commuting
`charges', with the hamiltonians $\hH_s$ arising from the single-particle
hamiltonian $h(p) = p^s$ forming a basis for this space. All such hamiltonians
have the same set of eigenstates, which are, then, independent of $h$ or $s$.

We shall, therefore, consider such a hamiltonian $\hH_s$, and, for simplicity,
will also subtract the zero-mode contribution to the energy:
\begin{align}\label{Htrue}
\hH_s &=\int p^s \,\hr(x,p)\,\dd x\,\dd p - E_0\,,\\
E_0&=\int p^s\,\vt(\hR_0-p)\,\dd p\,. \nonumber
\end{align}

A more manageable formula for $\hH_s$ can be obtained performing the sum over
$l$ and the $p$-integral in Eq.~\eqref{Htrue}, yielding
\begin{align}\nonumber
\hH&=\sum_{l=-\infty}^\infty\int\frac{\E^a\E^b}{16\pi^3\I\sin\frac k2}\,\E^{\I
k(\hR_0-p)-\I l(x-x')}\,\dd x\,\dd x'\,\dd k\,\dd p-E_0\\\nonumber
&=\sum_{l=-\infty}^\infty\sum_{i,j=0}^\infty\int\frac{p^s\,\E^{\I
k(\frac{N}2-p)+\I
ly}}{8\pi^2\I\, i!j!\,\sin\frac k2}a^ib^j\,\dd x\,\dd y\,\dd k\,\dd p-E_0\\
&=(-\I)^s\frac1{4\pi\I}\frac{\pd^s}{\pd k^s}\bigg|_{k=0}\bigg[\frac{\E^{\I
\frac{Nk}2}}{\sin\frac
k2}\int\bigg(\sum_{i,j=0}^\infty\frac{a^ib^j}{i!j!}-1\bigg)\,\dd x\bigg]\,.
\label{Hseries}
\end{align}
Despite the presence of a sine in the denominator, it is not difficult to check
that the integrand is in fact nonsingular at $k=0$.

Now we shall compute the lowest energy levels of this Hamiltonian. From
Eqs.~\eqref{a} and~\eqref{Hseries} it stems that $\hH$ preserves the subspace
of states of a given order because of the $x$-integral. (In fact, the order of
the state is simply the eigenvalue of the $\hH_1$ operator.) One should also notice
that the action of a term $\int a^ib^j\,\dd x$ over a state of order $r$
vanishes whenever $i$ or $j$ are greater than $r$. As a consequence of these
observations, $\ket0$ and $\ket1$ must be eigenfunctions of $\hH_s$ and $\ket0$
lies in its kernel.

The above observation also implies that the sole summand in Eq.~\eqref{Hseries}
whose action on the order-one state can be nonzero is
\[
\int\dd x\,ab\,\ket1=-8\pi\sin^2\tfrac k2\ket1\,.
\]
This easily leads to the formula
\begin{align*}
E_1&=2(-\I)^{s-1}\frac{\pd^s}{\pd k^s}\bigg|_{k=0}\big(\sin
\tfrac k2\,\E^{\I\frac{Nk}2}\big)\\
&=2^{1-s}\sum_{r=1}^{\lfloor\frac s2\rfloor}\binom
s{2r-1}N^{s-2r-1}\\
&=2^{-s}\big[(N+1)^s-(N-1)^s\big]
\end{align*}
for the first excited energy level. This result coincides with the exact
energy~\eqref{Eex1}.

It can be verified that higher energy levels are also exactly recovered. For
simplicity we shall concentrate on the second excitations of the system, which
can be obtained by diagonalizing the action of the Hamiltonian on the linear
span of $\ket2$ and $\ket{11}$. We shall prove that they do coincide with the
exact energies~\eqref{Eex2}.

By the preceding remark, the summands in Eq.~\eqref{Hseries} with $i,j>2$ yield
a vanishing contribution. In fact, a short calculation using the commutation
relations~\eqref{R_n} shows that the only terms which contribute to the
integral are
\begin{subequations}\label{aibj}
\begin{align}
\int\dd x\,ab\,\ket2&=-4\pi\sin^2k\,\ket2\,,\\
\int\dd x\frac{a^2b}2\,\ket2&=-8\pi\I\sin^2\tfrac k2\,\sin k\,\ket{11}\,,\\
\int\dd x\,ab\,\ket{11}&=-16\pi\sin^2\tfrac k2\,\ket{11}\,,\\
\int\dd x\,\frac{a^2b^2}4\,\ket{11}&=16\pi\sin^4\tfrac k2\,\ket{11}\,,\\
\int\dd x\,\frac{ab^2}2\,\ket{11}&=-8\pi\I\sin^2\tfrac k2\,\sin k\,\ket2\,.
\end{align}
\end{subequations}
Using Eqs.~\eqref{Hseries} and~\eqref{aibj} one can write the action of the
Hamiltonian on the state $\ket2$ as
\begin{align}\nonumber
\hH\ket2&=(-\I)^s\frac{\pd^s}{\pd
k^s}\bigg|_{k=0}\E^{\I\frac{Nk}2}\bigg(\frac{\I\sin^2k}{\sin\frac
k2}\,\ket2-2\sin\tfrac k2\,\sin k\,\ket{11}\bigg)\\\label{H2}
&=2^{-s}\sum_{r=0}^{\lfloor\frac s2\rfloor}\binom
s{2r-1}N^{s-2r+1}(1+3^{2r-1})\,\ket2+2^{-s}\sum_{r=0}^{\lfloor\frac s2\rfloor}
\binom s{2r}N^{s-2r}(3^{2r}-1)\,\ket{11}\,.
\end{align}

The analogous expression
\[
\hH\ket{11}=2^{-s}\sum_{r=0}^{\lfloor\frac s2\rfloor}\binom
s{2r-1}N^{s-2r+1}(1+3^{2r-1})\,\ket{11}+2^{-s}\sum_{r=0}^{\lfloor\frac
s2\rfloor} \binom s{2r}N^{s-2r}(3^{2r}-1)\,\ket{2}\,,
\]
can be obtained directly from the above equations using the elementary identity
$\sin\frac k2-\sin^3\frac k2=\frac{\sin^2k}{4\sin\frac k2}$. Therefore the
action of $\hH$ is diagonal in the basis $\frac1{\sqrt2}(\ket2\pm\ket{11})$ and
the corresponding eigenvalues are
\begin{align}\nonumber
E_{2,\pm}&=2^{-s}\sum_{r=0}^{\lfloor\frac s2\rfloor}\binom
s{2r-1}N^{s-2r+1}(1+3^{2r-1})\pm2^{-s}\sum_{r=0}^{\lfloor\frac s2\rfloor}
\binom s{2r}N^{s-2r}(3^{2r}-1)\\ \nonumber
&=2^{-s}\sum_{r=1}^s\binom srN^{s-r}\big[\pm(\pm3)^r\mp(\mp1)^r\big]\\
&=2^{-s}\big[\pm(N\pm3)^s\mp(N\mp1)^s\big]\,.\label{E2}
\end{align}
Thus the exact result~\eqref{Eex2} is reobtained, as we wanted to prove.

Notice that the eigenstates corresponding to the eigenvalues~\eqref{E2} just
obtained are, up to a normalizing factor, $\ket2\pm\ket{11}$. The coefficients
are not incidental: they can, in fact, be read off from the irreducible
representations of the symmetric group. Higher excited states of order $n$ can
similarly be constructed as characters of representations with $n$ boxes in
their Young tableaux. A complete proof of this statement is possible but will
not be given here.

The simplest examples of physical systems of the form~\eqref{Hsp} are those with
linear ($s=1$) and quadratic ($s=2$) dispersion relations. The $s=1$ system
can be interpreted as a harmonic oscillator in the action-angle basis, or as
a massless relativistic particle on a circle. The $s=2$ system would describe a
free nonrelativistic particle on a circle.

Interestingly, Eq.~\eqref{E_vr} shows that the semiclassical approximation is
{\it exact} for these systems. This is related to the fact that the expression
for $\hH_s$ obtained from (\ref{Htrue}) is of the form \beq \hH_s =
\frac{1}{(s+1)!} \int : \hR (x)^{s+1} :\dd x~+~{\rm lower~order~terms} \eeq
where the extra terms include powers $:\hR (x)^{s-1}:$ or lower, as well as
terms with derivatives of $\hR (x)$. Such terms are all trivial for $s=1,2$ and
therefore the semiclassical result is essentially exact.

\section{Conclusions}

We have derived the droplet expression for the fermion phase space density
satisfying the proper operator algebra. We also verified that it reproduces the
correct many-body fermion energies for non-interacting hamiltonians and that
it leads to the standard relativistic boson field theory and cubic collective
field theory for relativistic and nonrelativistic free fermions respectively.

The main thrust of this work, of course, is the higher dimensional situation.
For this, the corresponding expression for the higher dimensional phase space
density should be obtained. Again, a na\"\i ve (normal ordered) extension of
the corresponding semiclassical expression is not appropriate and a nontrivial
quantum modification would have to be performed.

This problem is presumably related, although not equivalent in any obvious way,
to the problem of identifying the Fermi operator. In ordinary (abelian)
one-dimensional bosonization this is given by the exponential of the boundary
droplet field. This expression is not directly motivated by any droplet
considerations, and does not generalize to the nonabelian case. Similarly, its
generalization to higher dimensions is an open issue. This and other questions
are the subject of further investigation.

\vskip0.5cm \noindent \underline{Achnowledgements:} We would like to thank
Dimitra Karabali for discussions. A.E.~acknowledges the financial support of
the Spanish Ministry of Education through an FPU scholarship and the partial
support of the DGI under grant no.~FIS2005-00752. A.P.'s research was supported
by NSF grant PHYS-0353301 and by RF-CUNY grants 67526-0036 and 80209-1312.
A.P.\ also acknowledges the
hospitality of the Aspen Center for Physics, where part of this work was done.


\begin{thebibliography}{99}\frenchspacing

\bibitem{Early}
F.~Bloch, Z.\ Phys.\ {\bf 81}, 363 (1933);
S.~Tomonaga, Prog.\ Theor.\ Phys.\ {\bf 5}, 544 (1950).

\bibitem{Bozo}
W.~Thirring, Ann.\ Phys.\ (N.Y.) {\bf 3}, 91 (1958);
J.~M.~Luttinger, J.\ Math.\ Phys.\ {\bf 4}, 1154 (1963);
D.\ Mattis and E.\ Lieb, J.\ Math.\ Phys.\ {\bf 6}, 304 (1965);
S.~R.~Coleman, Phys.\ Rev.\ D {\bf 11}, 2088 (1975);
S.~Mandelstam, Phys.\ Rept.\  {\bf 23} (1976) 307.

\bibitem{WZW}
E.~Witten,
Commun.\ Math.\ Phys.\  {\bf 92}, 455 (1984).

\bibitem{droplet}
J.~Polchinski, Nucl.\ Phys.\ B {\bf 362}, 125 (1991);
S.~Iso, D.~Karabali and B.~Sakita,
Phys.\ Lett.\ B {\bf 296}, 143 (1992) [arXiv:hep-th/9209003].
B.~Sakita, Phys.\ Lett.\ B {\bf 387}, 118 (1996);
D.~Karabali and V.~P.~Nair,
Nucl.\ Phys.\ B {\bf 641}, 533 (2002);
{\bf 679}, 427 (2004);
{\bf 697}, 513 (2004).

\bibitem{ncboz}
A.P.~Polychronakos,
``Bosonization in higher dimensions via noncommutative field theory",
Phys.\ Rev.\ Lett., in press [arXiv:hep-th/0502150]

\bibitem{hiboz}
A.~Luther, Phys.\ Rev.\ B  {\bf 19}, 320 (1979);
F.~D.~M.~Haldane, Varenna 1992 Lectures and cond-mat/0505529;
A.~Houghton and J.~B.~Marston, Phys.\ Rev.\ B {\bf 48}, 7790 (1993);
A.~H.~Castro Neto and E.~Fradkin, Phys.\ Rev.\ Lett.\ {\bf 72}, 1393 (1994);
D.~Schmeltzer and A.R.~Bishop, Phys.\ Rev.\ B {\bf 50}, 12733 (1994);
D.~V.~Khveshchenko, Phys.\ Rev.\ B {\bf 52}, 4833 (1995);
D.~Schmeltzer, Phys.\ Rev.\ {\bf 54}, 10269 (1996).


\bibitem{collect}
A.~Jevicki and B.~Sakita,
Nucl.\ Phys.\ B {\bf 165}, 511 (1980);
D.~Karabali and B.~Sakita,
Int.\ J.\ Mod.\ Phys.\ A {\bf 6}, 5079 (1991);
S.~R.~Das, A.~Dhar, G.~Mandal and S.~R.~Wadia,
Mod.\ Phys.\ Lett.\ A {\bf 7}, 71 (1992) [arXiv:hep-th/9111021]
Int.\ J.\ Mod.\ Phys.\ A {\bf 8}, 325 (1993) [arXiv:hep-th/9204028];
A.~Dhar, G.~Mandal and N.~V.~Suryanarayana,
JHEP {\bf 0601}, 118 (2006) [arXiv:hep-th/0509164].

\bibitem{phase}
A.~P.~Polychronakos,
Nucl.\ Phys.\ B {\bf 705}, 457 (2005) [arXiv:hep-th/0408194];
Nucl.\ Phys.\ B {\bf 711}, 505 (2005) [arXiv:hep-th/0411065].

\bibitem{sina}
D.~B.~Fairlie and C.~K.~Zachos,
Phys.\ Lett.\ B {\bf 224}, 101 (1989).

\bibitem{Wa}
I.~Bakas,
Phys.\ Lett.\ B {\bf 228}, 57 (1989);
A.~Cappelli, C.~A.~Trugenberger and G.~R.~Zemba,
Nucl.\ Phys.\ B {\bf 396}, 465 (1993) [arXiv:hep-th/9206027].

\bibitem{LLL}
S.~M.~Girvin, A.~H.~MacDonald and P.~M.~Platzman,
Phys.\ Rev.\ B{\bf 33}, 2481 (1986).

\bibitem{Dim}
D.~Karabali, private communication.




\end{thebibliography}
\end{document}